\documentclass[aps,pre,reprint]{revtex4-2}

\usepackage{amsmath}
\usepackage{amssymb}
\usepackage{graphicx}
\usepackage[separate-uncertainty,per-mode=symbol]{siunitx}
\usepackage{booktabs}

\makeatletter
\let\old@makecaption=\@makecaption
\usepackage[labelformat=simple]{subcaption}
\let\@makecaption=\old@makecaption
\makeatother

\renewcommand{\vec}[1]{\boldsymbol{#1}}
\renewcommand{\tensor}[1]{\underline{#1}}

\newcommand{\avg}[1]{\left< #1 \right>}
\newcommand{\real}[1]{\mathrm{Re}\left\{ #1 \right\}}
\newcommand{\imaginary}[1]{\mathrm{Im}\left\{ #1 \right\}}
\newcommand{\order}[1]{\mathcal{O}\left\{ #1 \right\}}
\newcommand\norm[1]{\left\lVert#1\right\rVert}

\begin{document}

\title{Nonreciprocal wave-mediated interactions power a classical time crystal}

\author{Mia C. Morrell*}

\author{Leela Elliott*}

\author{David G. Grier}

\affiliation{Department of Physics and Center for Soft Matter Research,
New York University, New York, New York 10003, USA}

\date{\today}

\begin{abstract}
An acoustic standing wave acts as a lattice of evenly spaced
potential energy wells for sub-wavelength-scale objects.
Trapped particles interact with each other
by exchanging waves that they scatter from the standing wave.
Unless the particles have identical scattering
properties, their wave-mediated interactions
are nonreciprocal.
Pairs of particles can use this nonreciprocity to harvest energy from
the wave to sustain steady-state oscillations
despite viscous drag
and the absence of periodic driving.
We show in theory and experiment that a minimal system
composed of two acoustically levitated particles
can access four distinct dynamical states,
two of which are emergently active
steady states.
Under some circumstances, these emergently active steady states break
spatiotemporal symmetry and therefore constitute
a classical
time crystal.
\end{abstract}

\maketitle

\begin{figure*}
    \centering
    \includegraphics[width=0.95\linewidth]{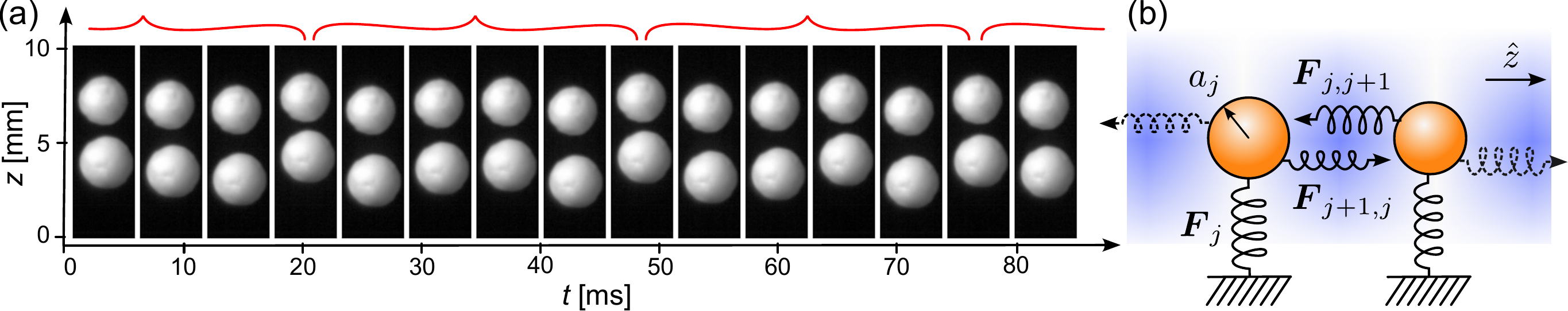}
    \caption{(a) Experimental realization of a steady-state time crystal composed of two millimeter-scale spheres of expanded polystyrene levitated in air by an acoustic
    standing wave at \SI{40}{\kilo\hertz}.
    Images captured at \SI{170}{frames\per\second} reveal
    sustained oscillations without periodic driving and despite dissipation due to viscous drag.
    Curly braces denote twice the \qty{16}{\milli\second} predicted period of
    the \qty{61}{\hertz} antisymmetric mode.
    (b) Model for the forces acting
    on spheres localized at the nodes of
    an acoustic standing wave,
    including restoring forces,
    $\vec{F}_j$, and nonreciprocal
     interparticle forces,
    $\vec{F}_{ij}$ and $\vec{F}_{ji}$.
    Dashed springs represent the possibility of extending
    the system to more than two particles.
    }
    \label{fig:experimental}
\end{figure*}

Waves exert forces on scatterers.
Scattered waves coalesce collections of independent scatterers
into self-organizing dynamical systems by mediating
interparticle interactions
\cite{burns1989optical,burns1990optical,marston1994bubble,rabaud2011acoustically,doinikov1996bubble,lim2022mechanical,lim2024acoustic}.
Wave-mediated pair interactions are not constrained
by Newton's third law because
the scatterers are not a closed system:
scattered waves can carry
away momentum, causing pairs of
interacting particles to recoil \cite{sukhov2015actio,sukhov2017nonconservative,raskatla2024continuous,king2025scattered}.
The nonreciprocity of wave-mediated interactions
allows scatterers to capture energy from the wave
and to transduce it into collective motion,
thereby endowing an otherwise passive system
with the defining characteristics of active matter
\cite{ramaswamy2010mechanics,marchetti2013hydrodynamics,bechinger2016active,shankar2022topological,vrugt2025exactlyactivematter}.
This form of activity is distinctive
because it is not an inherent property of the individual
particles, but instead is an emergent property of the
particles' configuration
\cite{king2025scattered}.

Here we demonstrate that emergent activity
can take the form of sustained oscillations
in an acoustically levitated array of particles.
Nonreciprocal wave-mediated interactions
continuously transfer energy from the levitator's standing wave into
the lattice's normal modes
without requiring periodic driving.
We show both in theory and through experiments that
a minimal system of two acoustically levitated
particles can
use this mechanism to access a variety of dynamical states,
including actively sustained steady states
\cite{sompolinsky1986temporal,mandal2024learning,weis2022coalescence},
one of which is a continuous classical time crystal
\cite{zaletel2023colloquium,raskatla2024continuous, liu2023photonic}.
Figure~\ref{fig:experimental}(a)
shows such a time crystal in action.

A classical time crystal is a dynamical
system whose
spatiotemporal symmetry is spontaneously broken.
The time crystal restores discrete spatiotemporal
symmetry through the emergence of a linearly
stable limit cycle \cite{raskatla2024continuous}.
Such a time crystal is said to be
continuous if the system's governing
equations are invariant under
continuous time translation,
in which case its oscillation frequency
is an emergent property.
This contrasts with discrete time crystals
that oscillate at harmonics or subharmonics
of an imposed driving frequency \cite{yao2020classical}.

The levitator used for this study is based on the popular TinyLev2
design \cite{marzo2017tinylev}, which operates at
\SI{40}{\kilo\hertz} in air and creates a linear array
of pressure nodes with a lattice constant of
\SI{4.3}{\mm}.
Each node acts as a three-dimensional potential energy
well for a millimeter-scale bead \cite{abdelaziz2020acoustokinetics,morrell2023acoustodynamic}
and exerts a nearly Hookean restoring force whose
stiffness depends on the bead's properties.
Trapped beads interact by exchanging
scattered waves, as depicted schematically
in Fig.~\ref{fig:experimental}(b).
Their resulting motions are recorded with a
video camera (Blackfly S, Teledyne Flir)
outfitted with a \qty{35}{\mm} $f$/1.4 lens
(Nikkor-n, Nikon) yielding a magnification
of \qty{139(1)}{\um\per pixel}.

The standing pressure wave along the
levitator's axis,
\begin{equation}
    p(z, t)
    =
    p_0 \sin(k z) \cos(\omega t),
\end{equation}
is characterized by its amplitude, $p_0$,
its frequency, $\omega$,
and its wave number,
$k = \omega / c_0$, in a medium of sound speed $c_0$.
Objects scattering this wave experience time-averaged
forces that organize them into a periodic lattice
along the wave's axis, $\hat{z}$, even if they
have different wave-scattering characteristics.
Heterogeneity in the trapped particles' properties constitutes
quenched disorder in the levitated array and
creates a context for nonreciprocal dynamics
\cite{king2025scattered}
that has not been addressed previously.

A sphere labeled $j$ and located
at $z_j$ within the standing wave
experiences a time-averaged force
originally formulated by Gor'kov
\cite{yosioka1955acoustic,gorkov1962forces,doinikov1994sphere},
\begin{subequations}
\label{eq:gorkov}
\begin{equation}
   \vec{F}_{j}(k z_j)
   =
    - \frac{\pi}{3} F_0 \,
    A_j x_j^3 \,
    \sin(2k z_j) \, \hat{z} ,
\end{equation}
whose scale,
\begin{equation}
    F_0
    =
    \frac{p_0^2}{\rho_0 \omega^2},
\end{equation}
is proportional to the
acoustic energy density in a medium of
mass density $\rho_0$.
Our levitator's calibrated
\cite{morrell2023acoustodynamic}
force scale is $F_0 = \SI{25.2(3)}{\micro\newton}$.
The Gor'kov force on a sphere also depends on the sphere's radius, $a_j$, relative
to the wavelength of sound,
\begin{equation}
    x_j = ka_j,
\end{equation}
as well as its density, $\rho_j$, and the ratio of
its isentropic compressibility, $\kappa_j$,
to that of the
medium, $\kappa_0$.
These material properties contribute to a
dimensionless coupling constant
\cite{king2025scattered},
\begin{equation}
    A_j = f_{0,j} + f_{1,j} ,
\end{equation}
\end{subequations}
that combines the spheres' monopole (pressure) and
dipole (velocity) polarizabilities,
\begin{subequations}
\begin{align}
    f_{0,j}
    & =
    1 - \frac{\kappa_j}{\kappa_0}
    \quad \text{and} \\
    f_{1,j}
    & =
    \frac{\rho_j - \rho_0}{2 \rho_j + \rho_0} ,
\end{align}
respectively.
\end{subequations}
Rigid spheres ($\kappa_j \ll \kappa_0$)
that are denser than the medium ($\rho_j > \rho_0$)
tend to be stably trapped at
nodes of the pressure field, located at
$k z_j = j \pi$.
This is the case for the  expanded polystyrene (EPS) beads
depicted in Fig.~\ref{fig:experimental}(a) that are
sufficiently incompressible relative to air
that we set $f_{0,j} \approx 1$.
The beads' mass density is
$\rho_j = \SI{30.5(2)}{\kilogram\per\cubic\meter}$
\cite{morrell2023acoustodynamic,horvath1994expanded},
relative to $\rho_0 = \qty{1.204(5)}{\kilogram\per\cubic\meter}$ for air, from which we obtain $f_{1,j} = \num{0.471(2)}$.

The wave-mediated interaction force
was originally formulated by K\"onig
in 1893
\cite{konig1891hydrodynamisch}
for the special case of identical
spheres in a common nodal plane.
This result recently has been generalized
to accommodate dissimilar spheres at
arbitrary positions within the wave \cite{king2025scattered}.
The leading-order generalized K\"onig
force on
sphere $j$ due to the wave scattered by
sphere $i$,
\begin{subequations}
    \label{eq:konig}
\begin{equation}
    \vec{F}^K_{ji}(k\vec{r}_{ji})
    =
    - 2 \pi F_0 \,
    f_{1,i} f_{1,j} \, x_j^3 x_i^3 \,
    \Phi(k\vec{r}_{ji}) \,
    \hat{r}_{ji} ,
\end{equation}
depends on the spheres' separation,
$\vec{r}_{ji} = \vec{r}_j - \vec{r}_i$,
through the dimensionless geometric
factor, $\Phi(k\vec{r})$.
For the special case of
spheres trapped in nodes along the axis of a
standing wave, the separation dependence
reduces to
\begin{equation}
    \Phi(k \vec{r})
    =
    \frac{\cos(kz) + kz \, \sin(kz)}{(kz)^2},
\end{equation}
\end{subequations}
which depends only on the trapped spheres'
axial separation, $z$.
The analytic expression in Eq.~\eqref{eq:konig} for the axial K\"onig
force appears not to have been reported
previously and is one of the contributions of this work.
Whereas the transverse K\"onig interaction
falls off as $r^{-4}$
\cite{konig1891hydrodynamisch,settnes2012forces,abdelaziz2020acoustokinetics,king2025scattered}, the
axial K\"onig force
falls off as $r^{-2}$ and thus
is longer-ranged.

Equation~\eqref{eq:konig} is strictly valid
only in the Rayleigh regime, which pertains
to scatterers that are
smaller than the wavelength of sound,
$x_j < 1$.
Comparison with
numerical studies \cite{stclair2023dynamics}, however,
shows that Eq.~\eqref{eq:konig}
captures qualitative features of the
inter-particle force for larger spheres
in the range $1 \leq x_j < 3$.

While Equation~\eqref{eq:konig}
is reciprocal under exchange of the indices $i$ and $j$
\cite{zheng1995acoustic,silva2014acoustic},
incorporating quadrupolar and octupolar
scattering
at leading order in $x_j$
\cite{king2025scattered}
yields corrections, $\chi_{ji}$, to
the pair interaction,
\begin{subequations}
\label{eq:generalizedaxialkonig}
\begin{equation}
\label{eq:F21}
    \vec{F}_{ji}(kz)
    =
    \vec{F}^K_{ji}(kz) \,
    (1 - \chi_{ji}) ,
\end{equation}
that are nonreciprocal
unless spheres $i$ and $j$ are identical.
For spheres made of the same material,
these higher-order corrections reduce to
two contributions:
\begin{equation}
\label{eq:chi21K}
    \chi_{ji}
    =
    \frac{2}{5} \, \sigma_{ji}(2)
    +
    \frac{1}{10}\Delta_{ji}(2),
\end{equation}
that depend on the spheres'
sizes through
\begin{align}
    \sigma_{ji}(n)
    & = x_j^n + x_i^n \quad \text{and} \\
    \Delta_{ji}(n)
    & = x_j^n - x_i^n .
    \label{eq:delta}
\end{align}
\end{subequations}
The latter of these factors changes
sign under exchange of indices and
therefore
describes a nonreciprocal contribution
to the pair interaction.
More generally, Eq.~\eqref{eq:delta}
establishes that spheres
of different sizes
interact nonreciprocally, as depicted
schematically in Fig.~\ref{fig:experimental}(b).

\begin{figure}
    \centering
\includegraphics[width=0.9\columnwidth]{timecrystal13_clean.pdf}
    \caption{(a) Dynamical states for a pair of levitated spheres
    of radii $a_1$ and $a_2$.
    Dashed curve: Rayleigh limit: $ka_1, ka_2 \leq 1$.
    Solid curves: roots of the
    stability functions: $\Lambda(n) = 0$.
    Activity surpasses dissipation in the (yellow) region
    bounded by
    $\Lambda(1) = 0$ and $\Lambda(5) = 0$.
    Circles denote experimental conditions from Fig.~\ref{fig:experimental}(a)
    and Fig.~\ref{fig:powerspectrum}.
    Shading represents the frequency of the antisymmetric mode
    relative to the natural frequency.
    (b) Frequency and (c) growth rate for the symmetric (blue) and antisymmetric (red)
    modes for pairs with $a_1 = \qty{1}{\mm}$ ($ka_1 = \num{0.732}$).
    Activity powers spontaneous emergence of oscillations in the (yellow)
    region where $\real{\lambda} \ge 0$.
    The steady-state symmetric mode is an active oscillator.
    The steady-state antisymmetric mode is a time crystal.
    }
    \label{fig:frequencies}
\end{figure}

Combining the Gor'kov force from Eq.~\eqref{eq:gorkov} with the generalized
axial K\"onig interaction from Eq.~\eqref{eq:generalizedaxialkonig}
yields the equations of motion for
an array of acoustically levitated spheres
trapped at the nodes of a standing wave.
The dimensionless displacement of the $j$-th sphere
from its trap,
$\zeta_j(\tau) = kz_j(t) - j \pi$,
evolves in time
according to the system of coupled equations,
\begin{subequations}
\label{eq:eom}
\begin{align}
    \nu_j
    & \equiv
    \dot{\zeta}_j \\
    \dot{\nu}_j
    & =
    -\frac{1}{2}\sin(2 \zeta_j)
    -
    \Gamma_j \, \nu_j
    +
    \sum_{i \neq j}
    B_{ji} \,
    \Phi(kz_{ji}) ,
\end{align}
where $kz_{ji} = (j-i) \, \pi + \zeta_j - \zeta_i$.
\end{subequations}
Dots in Eq.~\eqref{eq:eom}
represent derivatives
with respect to the dimensionless
time, $\tau = \Omega_0 t$,
which is scaled by the
natural oscillation frequency,
\begin{equation}
\label{eq:omega0}
    \Omega_0
    =
    \sqrt{\frac{F_0 A_j k^4}{2 \rho_j}} .
\end{equation}
For the EPS beads in our system,
$\Omega_0 = \qty{418(2)}{\radian\per\second}$
or \qty{66.5(3)}{\hertz}.
The interparticle coupling,
\begin{equation}
    B_{ji}
    =
    -3 \frac{f_{1,j}^2}{A_j}
    \, x_i^3 \,
    (1 - \chi_{ji}) \,
    \hat{r}_{ji} \cdot \hat{z} ,
\end{equation}
is nonreciprocal if the spheres differ in size.
Equation~\eqref{eq:eom} is based on the time-averaged
Gor'kov and K\"onig forces and therefore
is time-invariant.
For simplicity, we model dissipation
by Stokes drag acting
on the individual spheres independently
in a medium of viscosity
$\eta_0$.
Its influence is characterized by a
dimensionless damping rate,
$\Gamma_j = 2 \epsilon \, x_j^{-2}$, where
\begin{equation}
    \epsilon
    =
    \frac{9}{4}
    \frac{k^2 \eta_0}{\rho_j \Omega_0} .
\end{equation}
The expression for $\Gamma_j$
can be modified to account for inertial
effects in the fluid medium
\cite{maxey1983equation,landau2013fluid,settnes2012forces,morrell2023acoustodynamic}
without affecting the functional form of
Eq.~\eqref{eq:eom}.

The normal modes
for a pair of acoustically levitated spheres
labeled 1 and 2 are obtained
from the Jacobian of the equations of motion, $\tensor{J}(\zeta_1, \zeta_2; \nu_1, \nu_2)$,
by solving the characteristic equation,
\begin{equation}
\label{eq:characteristicequation}
    \left\vert
    \tensor{J}(0, 0; \nu_1, \nu_2) - \lambda \tensor{I}
    \right\vert = 0 ,
\end{equation}
for the eigenvalues, $\lambda$,
at the fixed point for the Gor'kov force, $\zeta_1 = \zeta_2 = 0$.

The full analytic expressions for $\lambda$ are unwieldy.
For clarity, we treat drag as a perturbation \cite{hinch1991perturbation},
$\epsilon < 1$,
to obtain an approximate expression
for the symmetric mode's eigenvalues
in units of $\Omega_0$:
\begin{subequations}
\label{eq:eigenvalues}
\begin{equation}
    \label{eq:symmetric}
    \frac{\lambda^{(s)}_{\pm}}{\Omega_0}
    =
    \pm i
    \, - \,
    \epsilon \,
    \frac{\Lambda(1)}{\Lambda(3)}
    +
    \order{\epsilon^2} .
\end{equation}
Similarly, the antisymmetric mode's eigenvalues are
\begin{equation}
\label{eq:antisymmetric}
    \frac{\lambda^{(a)}_{\pm}}{\Omega_0}
    =
    \pm i\sqrt{1 - g \, \Lambda(3)} \\
    -
    \frac{\epsilon}{x_1^2 x_2^2}\frac{\Lambda(5)}{\Lambda(3)}
    +
    \order{\epsilon^2} .
\end{equation}
Equation~\eqref{eq:eigenvalues} is suitable for our experimental system
because the viscosity of air is
$\eta_0 = \qty{1.825e-5}{\pascal\second}$
\cite{rumble2023crc}, so that
$\epsilon = \num{1.73(1)e-3}$.

The particles' sizes determine the
nature of these solutions
through stability functions,
\begin{equation}
\label{eq:stability}
    \Lambda(n)
    =
    \left[
    1 - \frac{2}{5}
    \sigma_{12}(2)
    \right]
    \sigma_{12}(n)
    +
    \frac{1}{10} \Delta_{12}(2) \Delta_{12}(n) ,
\end{equation}
whose roots establish boundaries between
the system's different dynamical states.
The frequency of the antisymmetric mode
additionally depends on particle size through
$\Lambda(3)$, and on the particles' composition
through
\begin{equation}
\label{eq:sigma}
    g
    =
    3 \frac{\pi^2 - 2}{\pi^3}
    \frac{f_{1,j}^2}{A_j} ,
\end{equation}
\end{subequations}
which has no subscript because
the particles in our minimal model are
all assumed to be made of the same material.
For the EPS beads used in our experiments,
$g = \num{0.1148(1)}$.

The normal mode frequencies,
$\Omega^{(s,a)}_\pm =
\imaginary{ \lambda^{(s,a)}_\pm }$,
are not influenced by drag to leading order in $\epsilon$.
These modes nevertheless tend to be damped
because
viscous drag typically removes energy from the system
faster than nonreciprocal interactions can replace it.
The passive state is a stable fixed point
of the dynamics under these conditions.
This prediction is consistent with the experimental
observation that most randomly selected pairs of spheres
remain motionlessly trapped in the acoustic levitator.

Not all systems have overdamped normal modes.
For pairs that satisfy $\Lambda(1)/\Lambda(3) \leq 0$,
for example, the initial growth
rate of the the symmetric mode is non-negative,
$\real{ \lambda^{(s)}_\pm} \geq 0$, which means that
the system spontaneously breaks into common-mode oscillation.
Similarly, the antisymmetric mode grows when
$\Lambda(5)/\Lambda(3) \leq 0$, leading to
spontaneous breathing-mode oscillations.
Roots of the stability functions,
$\Lambda(1) = 0$ and $\Lambda(5) = 0$,
therefore bound the conditions
where the system is emergently active.
These boundaries are plotted as solid curves
in Fig.~\ref{fig:frequencies}(a).
Higher-order contributions to the damping \cite{morrell2023acoustodynamic}
that are not included in Eq.~\eqref{eq:eom}
can stabilize the growing modes, which then settle
into steady-state oscillation.
The active region vanishes for particles of the same size
($a_1 = a_2$), whose interactions are reciprocal.

Regions in Fig.~\ref{fig:frequencies}(a) are colored by the
frequency of the lower-frequency normal mode obtained from
the full solutions to Eq.~\eqref{eq:characteristicequation}.
The condition $\Lambda(3) = 0$ divides
systems whose symmetric mode has the higher frequency
(blue) from those whose antisymmetric
modes are higher-frequency (red).
The latter systems break both
parity and time-parity symmetry
without periodic driving, as described in Appendix~\ref{sec:spatiotemporalsymmetrybreaking},
and therefore
have the defining characteristics of
continuous time crystals.

Figures~\ref{fig:frequencies}(b) and \ref{fig:frequencies}(c) present the
normal-mode frequencies and growth rates, respectively,
obtained from Eq.~\eqref{eq:characteristicequation}
for a typical family
of experimental systems with $a_1 = \qty{1}{mm}$
($x_1 = \num{0.72}$).
These traces show how
the active symmetric mode abruptly
crosses over to
an active antisymmmetric mode at $\Lambda(3) = 0$.
We demonstrate in Appendix~\ref{sec:numericalsimulations} that the
marginally-active antisymmetric state at $\Lambda(5) = 0$
($a_2/a_1 = \num{2.1}$)
has a linearly stable limit cycle and therefore is a
true continuous time crystal.

\begin{figure}
    \centering
    \includegraphics[width=\columnwidth]{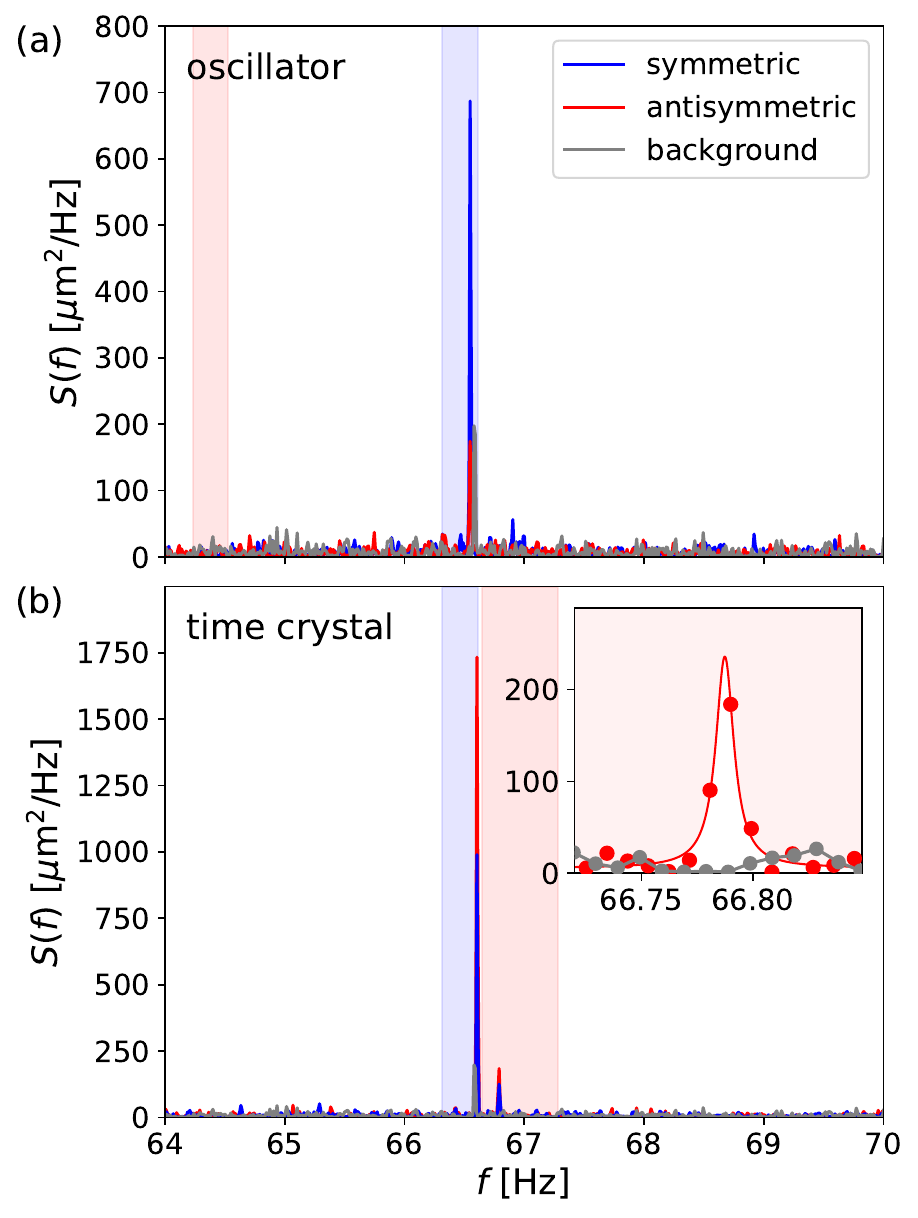}
    \caption{Power spectral density \cite{OSF}
    of the symmetric (blue) and antisymmetric (red) modes of
    two acoustically-coupled EPS beads
    in an acoustic levitator.
    Solid curves are computed from \num{20000}-frame
    video sequences acquired at \qty{200}{frames\per\second}.
    Shaded bands indicate the predicted
    frequency ranges for the normal
    modes from Eq.~\eqref{eq:eigenvalues}.
    Results for a single-particle
    trajectory (gray) represent
    the measurement's noise floor.
    (a) Small particles ($ka_1 = \num{0.80(2)}$ and $ka_2 = \num{1.02(2)}$) oscillate
    in the symmetric mode
    at the predicted common-mode
    frequency, $\Omega_0$.
    (b) Larger particles ($ka_1 = \num{1.07(4)}$ and $ka_2 = \num{1.21(3)}$) break
    spatiotemporal symmetry with
    antisymmetric oscillations at a higher
    frequency.
    Inset: Expanded view of the
    antisymmetric mode's spectral
    peak (red points) compared
    with a Lorentzian response
    (solid red curve) and the
    measured noise floor (gray points).
   }
    \label{fig:powerspectrum}
\end{figure}

The model defined by Eq.~\eqref{eq:eom}
provides a context for interpreting
observations of
steady-state oscillations in the experimental
system, such as the example in Fig.~\ref{fig:experimental}(a).
To facilitate this comparison,
the particles' measured positions in each
video frame are projected
along the axis of the acoustic levitator.
The common and relative modes are identified
from these time traces by principal component
analysis, and their frequency content is
assessed with the power spectral
density, $S(f)$, as described in Appendix~\ref{sec:experimentaltimeseriesanalysis}.
Examples computed from \qty{100}{\second} recordings
are plotted in
Fig.~\ref{fig:powerspectrum} for two different
pairs of spheres.
The software that implements this analytical pipeline is
available in the project's data repository \cite{OSF}.
Analogous results from other pairs of spheres and spot checks of the spheres in Fig.~\ref{fig:powerspectrum} are included in
Appendix~\ref{sec:additionaldatasets}.

The data in Fig.~\ref{fig:powerspectrum}(a)
are obtained with comparatively small spheres,
as indicated by the data point
in Fig.~\ref{fig:frequencies}(a),
and are predicted to be in the symmetric oscillator state.
The larger spheres that produced the data in Fig.~\ref{fig:powerspectrum}(b)
are predicted to form a time crystal.
Shaded regions in Fig.~\ref{fig:powerspectrum}
correspond to the ranges of frequencies
for the symmetric and antisymmetric modes
predicted by Eq.~\eqref{eq:eigenvalues}
after accounting for experimental
uncertainties.
The symmetric-mode frequency,
$\imaginary{\lambda^{(s)}_\pm} = \Omega_0$,
is predicted
to be independent of the particles' radii
and therefore should be the same for both systems
in Fig.~\ref{fig:powerspectrum}.
This is consistent with their observed behavior.
The antisymmetric mode of
the oscillator in Fig.~\ref{fig:experimental}(a)
is predicted to have a lower frequency,
$\imaginary{\lambda^{(a)}_\pm} < \Omega_0$, as drawn.
No spectral feature is observed at the predicted frequency,
however, presumably because the antisymmetric mode is
heavily damped.
The time crystal in Fig.~\ref{fig:powerspectrum}(b)
differs from
the oscillator by having
a higher antisymmetric-mode frequency,
$\imaginary{\lambda^{(a)}_\pm} > \Omega_0$.
Indeed, a small peak is observed
in the predicted frequency range in Fig.~\ref{fig:powerspectrum}(b).
The inset to Fig.~\ref{fig:powerspectrum}(b)
provides a magnified view of this peak
and compares it with a Lorentzian,
\begin{equation}
    S(f)
    =
    \frac{2\tau}{\pi} \frac{S_a}{1 + \left[2 (f - f_a) \, \tau \right]^2} + S_\text{noise},
\end{equation}
where $f_a = \qty{66.787(4)}{\hertz}$ is the antisymmetric mode's
center frequency,
$\tau = \qty{100(10)}{\second} = \qty{6700(700)}{cycles}$
is the antisymmetric mode's
coherence time,
and $S_a/S_\text{noise} = \num{40}$
serves as an estimate for the measurement's signal-to-noise ratio.
The estimated value of $\tau$ is comparable to the
\qty{108}{\second} duration of the data set and
exceeds the system's measured \cite{morrell2023acoustodynamic}
viscous relaxation time, \qty{0.16(1)}{\second},
by orders of magnitude.
Sustained oscillations persist for
hours longer than the measurement window,
which means that the true coherence time of
the nonequilibrium steady state may be substantially larger.
Comparable results are obtained for other pairs of
beads, as reported in Appendix~\ref{sec:additionaldatasets}.
Observed normal-mode frequencies fall into narrow
bands, which means that the results in Fig.~\ref{fig:powerspectrum} are typical.
This is consistent with the prediction that
activity precisely compensates dissipation
only over a
limited range of experimental conditions.

Consistency between the predicted and measured behavior of the levitated two-bead system
supports the contention that the experimental
system exhibits both active oscillator and
active time crystal states, and that both
display long-ranged temporal order.
This contrasts with previous reports of
dissipative time crystals \cite{kessler2021observation}
that require either
periodic driving \cite{taheri2022all} or
thermal activation \cite{yao2020classical}
to replace energy drained by dissipation.
The acoustically levitated system therefore
embodies a true classical time crystal
\cite{shapere2012classical,yao2020classical, taheri2022all}
that is powered by nonreciprocity and is
stabilized by damping
\cite{kessler2021observation,sacha2018time,taheri2022all}.

The equations of motion for this system, Eq.~\eqref{eq:eom}, also have
a fixed-point solution, $\nu_j = \dot{\nu}_j = 0$.
In this configuration, the forces acting on the spheres are
balanced and the spheres are motionless.
This is a passive state because
the system takes no energy from the
standing wave when the spheres are stationary.
The oscillatory steady states therefore are a manifestation
of emergent activity because the
system can
exist in active and passive states depending on its
configuration.
This contrasts with conventional active matter
that consumes energy continuously regardless
of its configuration.

Emergent activity also arises in larger arrays
of acoustically trapped particles.
In this context, quenched
disorder not only creates the conditions
required for sustained oscillations,
but also can mediate transitions between propagation and localization of linear excitations
\cite{sheng1990scattering}.
Analogous active steady-states should arise in any system where arrays of passive scatterers interact
with waves.
For example, scattering-mediated emergent activity
is likely to explain the classical time crystals recently observed in photonic metamaterials
\cite{liu2023photonic,raskatla2024continuous}
and the collective motions of buoyant particles
interacting with capillary waves \cite{ho2023capillary}.
Emergent activity therefore provides an unconventional
basis for designing compact oscillators,
resonant detectors and time bases for technological
applications.

\begin{acknowledgments}
  This work was supported by the National Science Foundation
  under Award Nos.~DMR-2104837 and
  DMR-2428983.
  The authors acknowledge an exceptionally fruitful exchange of ideas with the three anonymous reviewers in addition to
  helpful conversations with Vincenzo Vitelli,
  Paul Chaikin,
  Ella King, Mathias Casiulis,
  Andriy Goychuk,
  Ankit Vyas and
  Matthew Gronert.
\end{acknowledgments}


\section*{Appendices}

\appendix

\section{Material Properties}
\label{sec:materialproperties}

Table~\ref{tab:materials} reports material properties
for the experiments on acoustically levitated pairs of spheres
described in the main text. The same parameters are used in simulations.
Properties of air are obtained from Ref.~\cite{rumble2023crc}.
Reference~\cite{morrell2023acoustodynamic} describes
the methods used to measure the force scale, $F_0$, of the acoustic
trap and the density, $\rho_j$, of the levitated particles.
Equation numbers refer to definitions of derived quantities
in the main text.

\begin{table*}[ht]
    \centering
    \caption{Material properties of the experimental system.}
    \label{tab:material_params}
    \begin{tabular}{@{} l c S l @{}}
        \toprule
        \textbf{Parameter} & \textbf{Symbol} & \textbf{Value} & \textbf{Source} \\
        \midrule
        \multicolumn{4}{l}{\textbf{(a) Properties of sound wave}} \\
        speed of sound in air & $c_0$        & $\qty{343(1)}{\meter\per\second}$   & measured from node separation \\
        viscosity of air   & $\eta_0$        & \qty{1.825e-5}{\pascal\second}        & Ref.~\cite{rumble2023crc} \\
        acoustic frequency & $\omega/(2\pi)$ & \qty{40}{\kilo\hertz}                 & instrumental setting \\
        wave number        & $k$             & \qty{732(1)}{\radian\per\meter}       & computed \\
        air density        & $\rho_0$        & $\qty{1.204(5)}{\kg\per\cubic\meter}$ & Ref.~\cite{rumble2023crc} \\
        force scale        & $F_0$           & \qty{25.2(3)}{\micro\newton}               & Eq.~(2b) \\
        \midrule
        \multicolumn{4}{l}{\textbf{(b) Properties of particles}} \\
        particle density   & $\rho_j$        & \qty{30.5(2)}{\kilogram\per\cubic\meter} & Ref.~\cite{morrell2023acoustodynamic} \\
        pressure polarizability & $f_{0,j}$  & \num{1}                               & estimated \\
        velocity polarizability & $f_{1,j}$  & \num{0.471(2)}                        & Eq.~(3b) \\
        \midrule
        \multicolumn{4}{l}{\textbf{(c) Derived dynamical properties}} \\
        natural frequency  & $\Omega_0/(2\pi)$ & \qty{66.6(6)}{\hertz}                    & Eq.~(7) \\
        drag parameter     & $\epsilon$      & \num{1.72(1)e-3}                      & Eq.~(9) \\
        antisymmetric mode frequency factor  & $g$ & \num{0.1148(1)}                         & Eq.~(11d) \\
        \bottomrule
    \end{tabular}
    \label{tab:materials}
\end{table*}

\section{Spatiotemporal symmetry breaking and exceptional points}
\label{sec:spatiotemporalsymmetrybreaking}

\begin{figure*}
  \begin{minipage}[t]{0.48\textwidth}
    \centering
    \raisebox{-7in}{\includegraphics[height=7in]{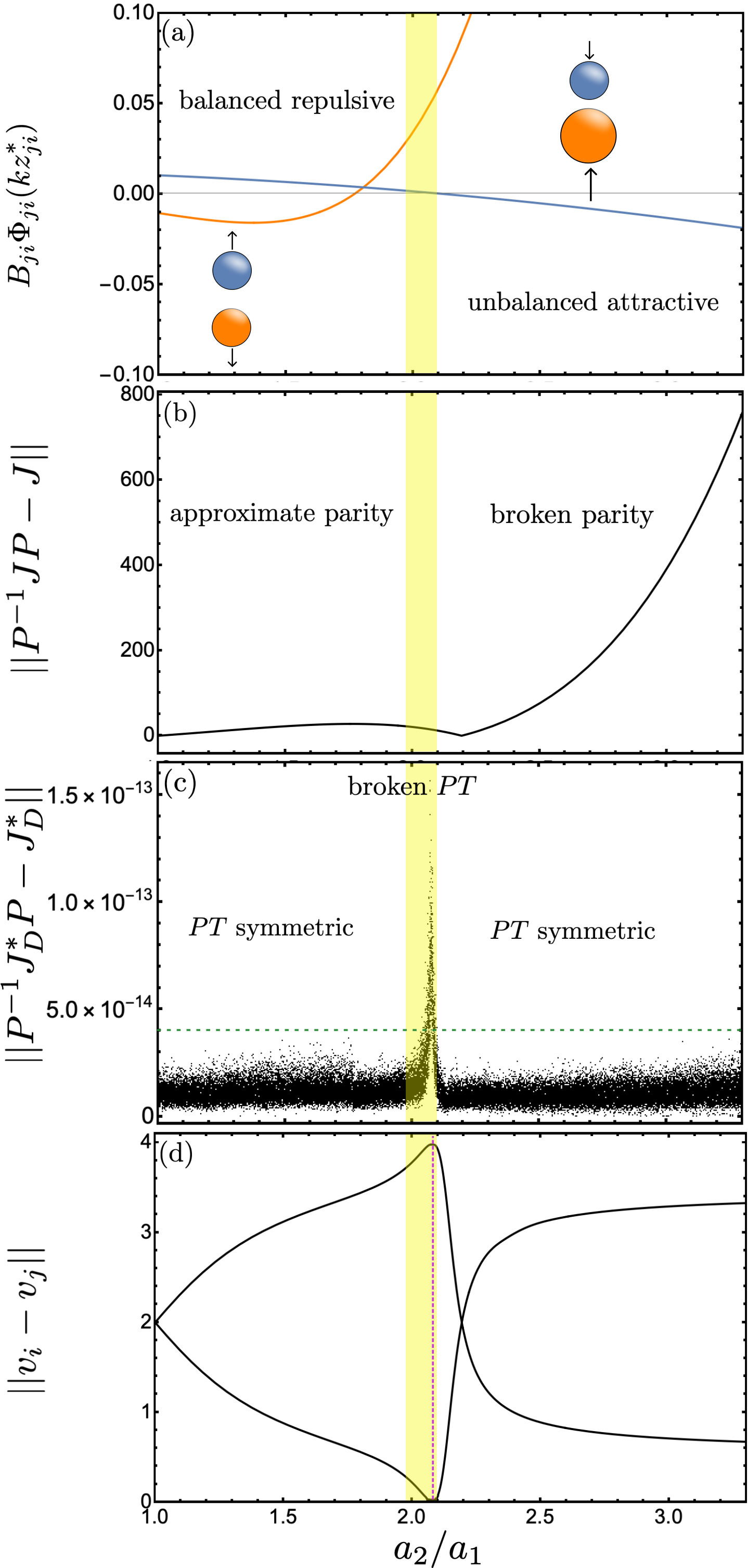}}
  \end{minipage}
  \hfill
  \begin{minipage}[t]{0.48\textwidth}
    \caption{(a) The fixed point interaction force between two particles of radii $a_1$ (blue curve) and $a_2$ (orange curve) as a function of the
    particle size ratio, $a_2/a_1$. When the particle size ratio is small, the interaction force between the particles is repulsive and approximately reciprocal. A large particle size ratio results in an attractive interparticle force that is \emph{nonreciprocal}: the force on particle 2 due to particle 1 is much stronger than the force on particle 1 due to particle 2.
    The yellow region corresponds to the active region in Fig.~2 in the main text, within which nonreciprocal interactions replace the energy lost to dissipation.
    (b) The Frobenius norm of the difference between the Jacobian for the levitated two-particle system and the parity-transformed Jacobian as a function of the particle size ratio.
    Parity is approximately conserved when the interaction force is repulsive.
 When the interaction force is attractive, parity is strongly broken.
 (c) The Frobenius norm of the difference between the Jacobian for the levitated two-particle system and the $PT$-transformed Jacobian as a function of the particle size ratio. Parity-time symmetry is broken within the (yellow) active range.
 The horizontal dotted line represents a noise floor below which all norm differences are indistinguishable from zero.
 (d) The norm difference between all unique pairs of the Jacobian's eigenvectors, $v_i$.
 An exceptional point occurs when the norm difference is zero, as marked by the vertical dotted line.
 The exceptional point occurs within the active range.}
\label{fig:symmetry}
\end{minipage}
\end{figure*}

The main text demonstrates that the antisymmetric-mode
frequency for a pair of acoustically levitated particles equals the symmetric-mode frequency
for certain particle-size ratios.
Specifically, small
particles ($ka_1 = \num{0.80(2)}$ and $ka_2 = \num{1.02(2)}$) oscillate
in the symmetric mode
at the predicted common-mode
frequency, $\Omega_0$, while larger particles ($ka_1 = \num{1.07(4)}$ and $ka_2 = \num{1.21(3)}$) break
spatiotemporal symmetry with
antisymmetric oscillations at a higher
frequency.
Here we show that the specific spatiotemporal symmetry broken
is parity-time ($PT$) symmetry.

\subsection{Parity symmetry}

A system with parity symmetry is one whose dynamics are unchanged upon reflection about a center of symmetry.
An example of such a reflection is the  exchange of particle identity.
Specifically, if a system's dynamics about a fixed point are given by the differential equation
\begin{equation}
    \dot{\vec{y}} = J \vec{y}
\end{equation}
with fixed-point Jacobian $J$, then approximate parity symmetry corresponds to the condition
\cite{shankar1994principles}
\begin{eqnarray}
    \norm{P^{-1}JP-J} \approx 0,
\end{eqnarray}
where $P$ is the operator that swaps particle identities.
Double bars denote the Frobenius norm of the enclosed matrix.

Systems with nonreciprocal interactions, such as a pair of acoustically levitated particles, break parity symmetry by definition.
While parity symmetry is always broken when the levitated particles are not identical, parity symmetry can be approximately conserved if $P$ and $J$ almost commute.
Two acoustically levitated particles, for example, approximately conserve parity symmetry if their size ratio is small ($a_2/a_1\leq 2$) as shown in Fig.~\ref{fig:symmetry}(a) and (b).
In Fig.~\ref{fig:symmetry}(a), we observe that the interaction force between two similarly sized particles is repulsive and approximately balanced: the force on particle 1 due to particle 2 is similar in magnitude to the force on particle 2 due to particle 1.
This approximate parity symmetry is explicitly observed in Fig.~\ref{fig:symmetry}(b), which shows the norm difference between the Jacobian and its parity-transform.

Alternatively, parity symmetry can be strongly broken if $P$ and $J$ do not commute.
In  Fig.~\ref{fig:symmetry}(a), for example, we observe that the interaction force between two very heterogeneous particles ($a_2/a_1> 2$) is attractive and unbalanced in magnitude.
We see this parity violation explicitly in Fig.~\ref{fig:symmetry}(b), where these
dissimilar particles strongly break parity symmetry.

\subsection{Parity-time symmetry and exceptional points}

While the conservation or violation of parity symmetry describes how energy is transferred between particles, that of parity-time ($PT)$ symmetry helps us understand how energy is allocated to our system's dynamical modes.
A $PT$  symmetric system is one whose dynamics are invariant under reflection across an axis of symmetry plus a time reversal.
A system with approximate $PT$ symmetry obeys
the condition \cite{shankar1994principles,Miri2019Exceptional}
\begin{eqnarray}
    \norm{P^{-1}J_{D}^*P-J_D} \approx 0,
\end{eqnarray}
where the parity operator $P$ performs a reflection about a center of symmetry and $J^*_{D}$ is the complex conjugate of the diagonalized Jacobian, $J_D$.
When $PT$ symmetry is broken it can be restored through a limit cycle \cite{fruchart2021non}.
If the frequency of this limit cycle is uncorrelated with timescales which appear in the system's equations of motion, the limit cycle realizes a time crystal.
Furthermore, in photonics and optics this spontaneously broken $PT$ symmetry can coincide with the coalescence of two eigenvectors, after which the Jacobian can no longer be diagonalized \cite{Miri2019Exceptional}.
These special points, known as ``exceptional points'', are typically found in non-Hermitian quantum mechanics \cite{ Miri2019Exceptional} and also can arise in nonreciprocal classical systems \cite{fruchart2021non}.

Our levitated two-particle system has two dynamical modes describing  symmetric and antisymmetric motion.
These dynamical modes consist of the two complex conjugate pairs of eigenvalues in main text's Eq.~\eqref{eq:eom} and their corresponding eigenvectors.
The relevant parity operator for our system swaps the identity of the eigenvalues within each complex conjugate pair.
As long as a unique antisymmetric and symmetric mode exist, $PT$ symmetry is conserved.
Accordingly, in Fig.~\ref{fig:symmetry}(c) we observe that before and after the bifurcation shown in Fig.~2(b) in the main text, the system of two acoustically levitated particles is approximately $PT$ symmetric.
At the bifurcation in Fig.~2(b) in the main text, however, the symmetric and antisymmetric eigenvalues coalesce and the system strongly breaks $PT$ symmetry.
This spontaneous spatiotemporal symmetry breaking, embodied by the peak in Fig.~\ref{fig:symmetry}(b), corresponds with the coalescence of two eigenvectors, as shown in Fig.~\ref{fig:symmetry}(d).
The $PT$ symmetry breaking shown in Fig.~2 therefore coincides with an exceptional point.
At this exceptional point, the broken $PT$ symmetry is restored by a limit cycle, which is the time crystal shown in Fig.~2 of the main text.
On this basis, we conclude that the steady-state
antisymmetric mode of the dynamical system
described by Eq.~\eqref{eq:eom} in the main text
is a continuous classical time crystal.

\section{Numerical simulations: Stability of the active dynamical states}
\label{sec:numericalsimulations}

Referring to Eq.~\eqref{eq:eom} of the main text,
we simulate the dynamics of a pair of acoustically levitated particles
by numerically integrating the coupled Langevin equations,
\begin{subequations}
\label{eq:neom}
\begin{align}
    \nu_j
    & \equiv
    \dot{\zeta}_j \\
    \dot{\nu}_j
    & =
    -\frac{1}{2}\sin(2 \zeta_j)
    -
    \Gamma_j \, \nu_j
    + \sum_{i \neq j} B_{ji} \,
    \Phi(\pi + \zeta_2 - \zeta_1)
    +
    \xi_j(t) ,
    \label{eq:acceleration}
\end{align}
\end{subequations}
with indices $j = 1, 2$.
The $j$-th particle is located at $\zeta_j$ relative
to the node of its acoustic trap in
units of the acoustic wavelength.
The associated dimensionless velocity is $\nu_j$.
Dots denote derivatives with respect to time
in units of the period of the acoustic traps'
natural frequency.
Each particle tends to be localized at a node
of the standing wave, $\zeta_j = 0$, by the
primary Gor'kov force.
The spheres' motions are damped by viscous drag
with dimensionless drag coefficients,
$\Gamma_j = 2 \epsilon (k a_j)^2$,
that depends on the spheres' radii, $a_j$, relative
to the wavelength of sound.
The magnitude of the drag force is
set by the dimensionless parameter $\epsilon < 1$,
which is expressed in terms of the system's physical
properties in the main text.

Wave-mediated interactions displace the particles from
the centers of their traps.
These scattering forces depend on the dimensionless
interparticle separation through
\begin{equation}
    \Phi(kz)
    =
    \frac{\cos(kz) + kz \, \sin(kz)}{(kz)^2} .
\end{equation}
The coupling constants,
\begin{eqnarray}
    B_{12}
    & =
    -3 \frac{f_{1,1}^2}{f_{0,1} + f_{1,1}} \,
    (k a_2)^3 \, (1 - \chi_{12}) \\
    B_{21}
    & =
     3 \frac{f_{1,2}^2}{f_{0,2} + f_{1,2}} \,
    (k a_1)^3 \, (1 - \chi_{21}) ,
\end{eqnarray}
incorporate the coefficients,
\begin{equation}
    \chi_{ij} = \frac{1}{2} (ka_i)^2
    \left(1 + \frac{3}{5} \frac{a_j^2}{a_i^2}\right)
\end{equation}
to order $(k a_j)^5$ in the multipole expansion for
the interparticle interaction.
This result for the axial K\"onig interaction
between dissimilar pairs of spheres is one of the original
contributions of this work.
The pair interaction is nonreciprocal when
$B_{12} \neq -B_{21}$, which is the case when
the particles are composed of the same material but have different radii.
The simulated particles also are subject to Gaussian random forces
with correlations
\begin{equation}
    \avg{\xi_i(t) \, \xi_j(0) }
    =
    2 \, s^2 \, \delta_{ij} \delta(t) ,
\end{equation}
that are characterized by a dimensionless amplitude of order
$s \approx \num{e-3}$.
This small amount of noise is intended to assess
the linear stability of the dynamical states
identified in the main text.
It is not intended to model thermal noise or
environmental perturbations.

\begin{figure*}
    \centering
    \includegraphics[width=0.65\textwidth]{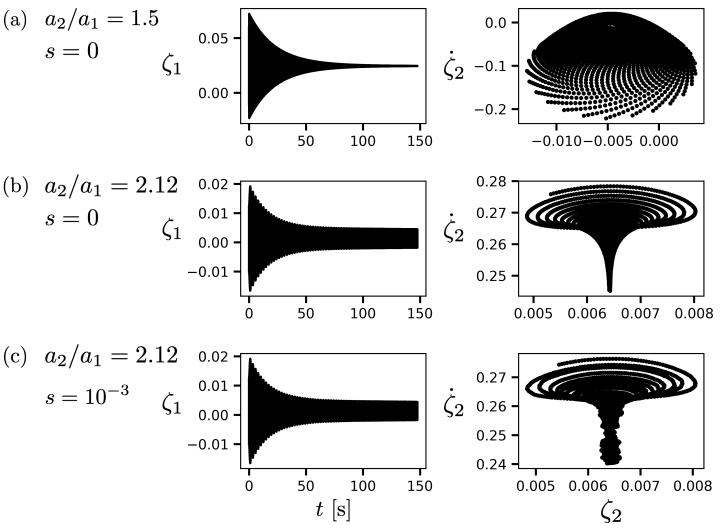}
    \caption{Numerical solutions and Poincar\'e maps of Eq.~\eqref{eq:neom} for two-particle systems with $a_1 = \qty{1}{\mm}$.
    (a) $a_2 = \qty{1.5}{\mm}$: In the absence of noise ($s = 0$),
    the system's oscillations ring down to a stable fixed point ($\dot{\zeta}_1 = \dot{\zeta}_2 = 0$).
    (b) $a_2 = \qty{2.12}{\mm}$, $s = 0$: Analytically predicted conditions for a true time crystal. The trajectory starts from rest and evolves toward a steady-state limit cycle
    corresponding to the true continuous time crystal.
    (c) $a_2 = \qty{2.12}{\mm}$, $s = \num{e-2}$:
    A small amount of additive force noise does not prevent
    the system from evolving toward the steady-state limit cycle,
    confirming that the time crystal is linearly stable.}
    \label{fig:poincare}
\end{figure*}

A simulation is configured with six dimensionless parameters:
$ka_1$ and $ka_2$, $B_{12}$, and $B_{21}$, the drag coefficient
$\epsilon$, and the noise coefficient $s$.
These parameters can be related to the physical properties
of an experimental system through
the values tabulated in Table~\ref{tab:materials}.
For comparison with the experiments described in the main text,
we choose $a_1 = \qty{1}{\mm}$ ($ka_1 = \num{0.732}$) and
$a_2 > a_1$.
We integrate this system of equations
with the fourth-order Runge-Kutta method
implemented in the \texttt{solve\_ivp} routine that is provided by \texttt{scipy.integrate} \cite{scipy}.
Typical trajectories and their Poincar\'e sections appear in Fig.~\ref{fig:poincare}.

In the absence of noise, the null vector,
$\nu_j = \dot{\nu}_j = 0$, is
a solution of Eq.~\eqref{eq:neom}
for any pair of spheres that can be stably trapped.
This fixed point corresponds to a
set of particle positions, $\zeta_j$ and $\zeta_i$,
that satisfy
\begin{equation}
    \frac{1}{2}\sin(2\zeta_j) = \sum_{i \neq j}
    B_{ji} \,
    \Phi(kz_{ji}).
    \label{eq:equilibrium}
\end{equation}
The system neither dissipates energy in this configuration
nor extracts energy from the standing wave.
Such a fixed point therefore is a passive state.

Figure~\ref{fig:poincare} presents numerically evaluated results
for a pair of acoustically levitated spheres under conditions
specified in Table~\ref{tab:simulation}.
The first column in Fig.~\ref{fig:poincare} shows the trajectory
of particle 1, $\zeta_1(t)$,
starting from an arbitrary displacement, $\zeta_1(0) = -0.01$ and $\zeta_2(0) = 0.01$.
The second column shows Poincar\'e sections of the two-particle trajectories
in the $\zeta_2$-$\dot{\zeta}_2$ plane obtained when particle 1 passes
through its equilibrium position
with $\dot{\zeta}_1 = 0$.

\begin{table*}[ht]
    \centering
    \caption{Parameters used in Fig.~\ref{fig:poincare}.
    Particle 1 radius: $a_1 = \qty{1}{\mm}$;
    particle 1 drag coefficient: $\Gamma_1 = \num{6.5e-3}$.}
    \begin{tabular}{@{} l S S S S S @{}}
    \toprule
    & $a_{2}$ & $\Gamma_{2}$ & $B_{12}$ & $B_{21}$ & $s$ \\
    \midrule
    Figure~\ref{fig:poincare}(a) & \qty{1.5}{\mm} & \num{2.9e-3} & \num{-2.2e-1} & \num{4.2e-2} & 0 \\
    Figure~\ref{fig:poincare}(b) & \qty{2.12}{\mm} & \num{1.5e-3} & \num{-1.3e-2} & \num{-6.6e-2} & 0 \\
    Figure~\ref{fig:poincare}(c) & \qty{2.12}{\mm} & \num{1.5e-3} & \num{-1.3e-2}& \num{-6.6e-2} & \num{e-3} \\
    \bottomrule
    \end{tabular}
    \label{tab:simulation}
\end{table*}

Figure~\ref{fig:poincare}(a) shows a typical oscillator state in which the
pair trajectory rings down to a static equilibrium. The particles' nonreciprocal
interactions do not provide enough energy to compensate for dissipation under
these conditions and the system tends toward its fixed point.
The fixed point is stable for such a system.

The active time-crystal state in Fig.~\ref{fig:poincare}(b) shows how a
two-particle system behaves when nonreciprocity precisely compensates for
drag. In this case, the system selects a steady-state amplitude and the
Poincar\'e map converges on a state in which particle 2 remains in motion.
Precise matching of activity and drag is achieved for particle sizes
satisfying $\Lambda(5) = 0$, as explained in the main text.
The prediction that such a system constitutes a true time crystal
is consistent with the results of this simulation.
The system also converges on the same steady-state time-crystal
solution for other
choices of the initial positions in the range $\zeta_1, \zeta_2 \in (-0.5, 0.5)$.

The continuous time-crystal persists even in the presence of random
force noise, as shown in Fig.~\ref{fig:poincare}(c).
This result incorporates a comparatively large amount of noise,
$s = \num{e-3}$.
The Poincar\'e map nevertheless converges into a region around
the same limit cycle observed in Fig.~\ref{fig:poincare}(b).
This demonstrates that the true time-crystal state is linearly stable
against perturbations.
Larger amplitudes of noise, $s \gtrsim \num{e-2}$, destabilize
the time crystal state and can disrupt the two-particle system
altogether by enabling one or both of the particles to escape their traps.

\section{Experimental time series analysis}
\label{sec:experimentaltimeseriesanalysis}

\begin{figure}[h]
    \centering
    \includegraphics[width=0.9 \columnwidth]{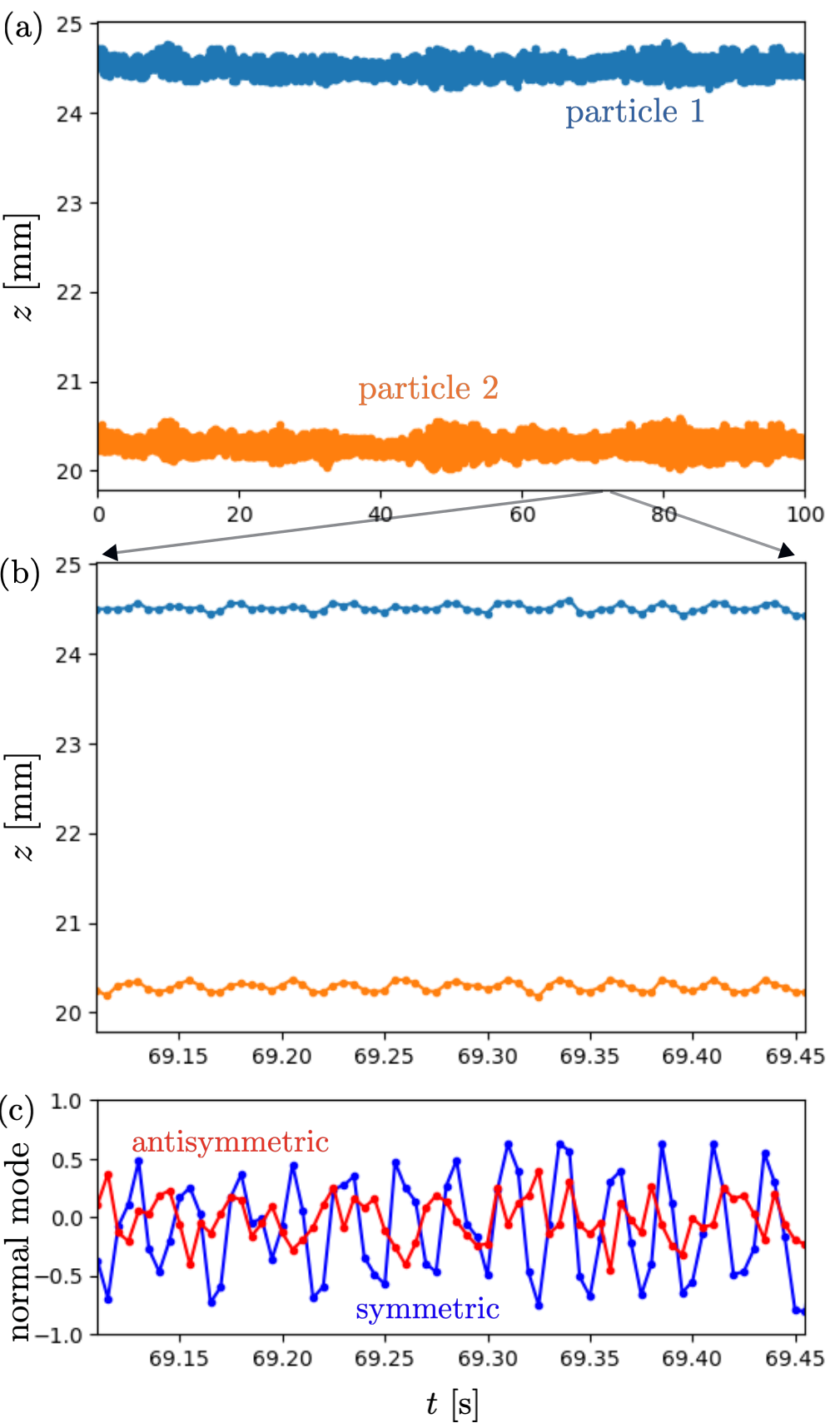}
    \caption{(a) Measured trajectories of the pair
    of levitated EPS beads presented in Fig.~3(b) of the main text. (b) Detail of the time series showing the particles' coherent oscillations. (c) Symmetric (blue) and antisymmetric (red) components of the trajectory data from (b). }
    \label{fig:traj}
\end{figure}

Data and analysis code used to create the figures in the main text are available from OSF repository \cite{OSF}.
Figure~\ref{fig:traj}(a) shows the measured trajectories,
$z_1(t)$ and $z_2(t)$,
for the two spheres in the steady-state time crystal
presented in Fig.~3(b) of the main text.
Each sphere is located to within \qty{0.3}{pixel} (\qty{40}{\um}) in each video frame
by thresholding the image
and finding the center of mass of each of the the simply-connected islands of pixels.
This analysis also yields estimates for the
particles' radii.
Figure~\ref{fig:traj}(b) shows a
\qty{300}{\ms} time window in which
the individual particles' fluctuations are
more clearly visible.

\begin{figure}
    \centering
    \includegraphics[width=0.9\columnwidth]{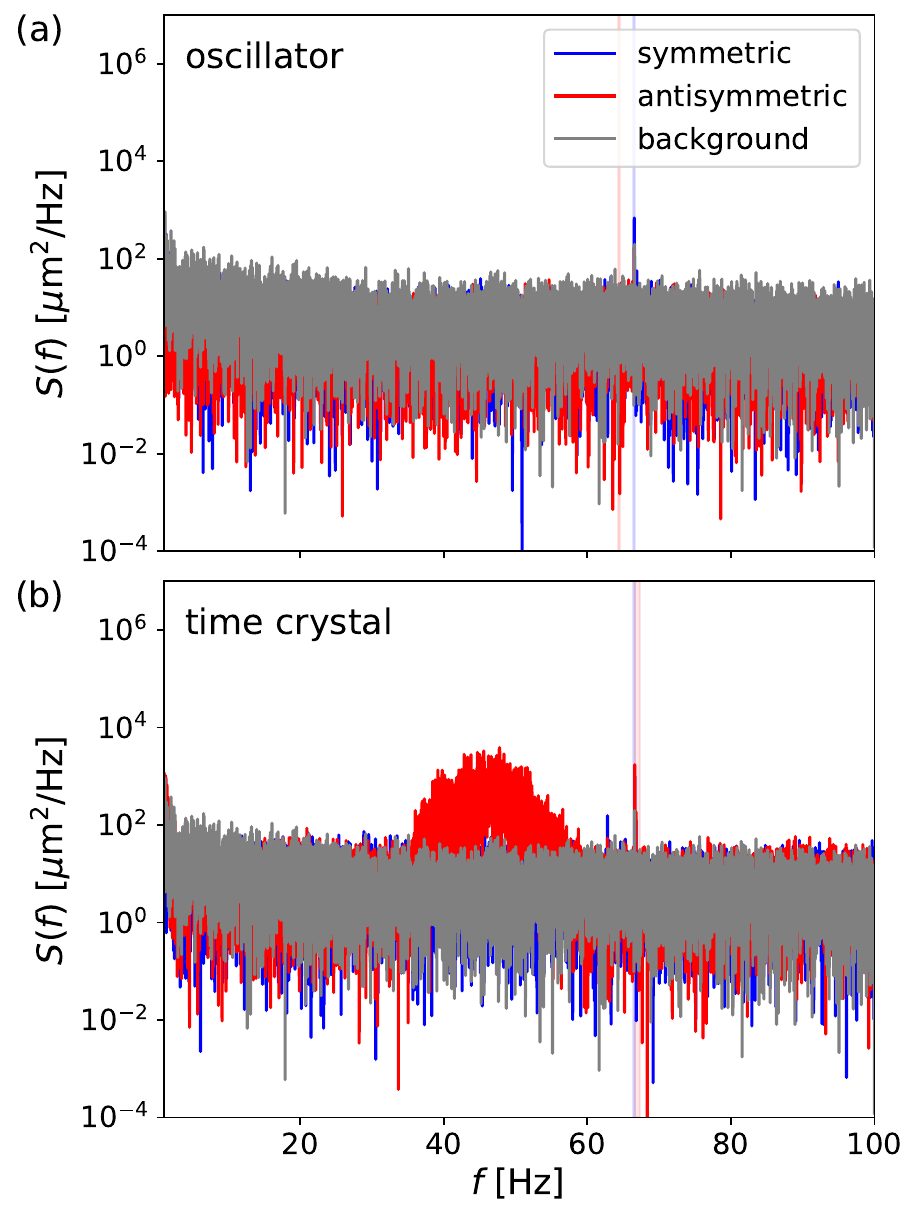}
    \caption{Power spectra for the pairs of particles reported in
    Fig.~3 of the main text, replotted on semilogarithmic axies
    over a larger frequency range. (a) Active oscillator
    showing strong oscillations at the predicted symmetric-mode
    frequency, plotted as a (blue) vertical shaded bar.
    No response is apparent at the frequency for the
    antisymmetric mode (red).
    (b) Steady-state time crystal displaying a spectral peak
    at the predicted frequency for antisymmetric oscillations (red).
    Large-amplitude in-line oscillations excite transverse oscillations that contribute to the broad low-frequency
    response in the symmetric mode (blue).}
    \label{fig:powerspectra}
\end{figure}

The pair's normal modes are estimated from
the full \num{20000}-step time series using
the python implementation of principal
component analysis provided by the
\texttt{sklearn.decomposition} library
of \texttt{scikit-learn} \cite{scikit-learn}.
Figure~\ref{fig:traj}(c) shows the result
of that decomposition over the same time window
presented in Fig.~\ref{fig:traj}(b).

The power spectral density for each
normal-mode signal is computed
using Welch's method \cite{welch2003use}
as implemented in \texttt{scipy.signal}
\cite{scipy}.
A typical time series consisting of
\num{20000} pair positions
acquired at \qty{200}{frames\per\second}
yields an effective frequency resolution
of \qty{0.01}{\hertz}.
Figure~\ref{fig:powerspectra} shows the
power spectral
densities from Fig.~3 of the main text replotted
on a semilogarithmic scale over a wider range of frequencies.
Blue traces show contributions to the symmetric modes obtained
from principal component analysis of the pair trajectories, and red traces show
contributions to the antisymmetric modes.
Vertical blue and red bars represent the predicted range of
normal-mode frequencies for symmetric and antisymmetric modes, respectively.
Gray traces in Fig.~\ref{fig:powerspectra} show the
power spectral densities for the trajectory of a single sphere
trapped in one node of the acoustic levitator.
This establishes the noise floor for the measurement
and is consistent with the observed background for the
two-particle normal modes.

In addition to sustained oscillations at the predicted
symmetric-mode and antisymmetric-mode frequencies, the
system presented in Fig.~\ref{fig:powerspectra}(b) also displays
strong broad-band spectral features at lower frequencies
in its common-mode motion (blue curve).
These low-frequency spectral features result from coupling
between the particles' axial and transverse degrees of freedom
arising from the overall large amplitude of these particles'
oscillations.
The particles' transverse displacements are not accounted for
by Eq.~\eqref{eq:eom}, which is a one-dimensional model.
The acoustic levitator’s traps are weaker in the transverse direction
than in the axial direction, which means that transverse displacements
occur at lower frequencies and with larger amplitudes than
coherent axial
oscillations.
The axial and transverse modes are linearly independent for
small-amplitude oscillations.
At larger amplitudes, however, the nonlinearity of the
trapping potential couples transverse displacements into
the axial degrees of freedom, leading to the observed
broad response.
Interpreting these observation in light of the one-dimensional
theory, we propose that the antisymmetric mode is actively
powered under the conditions of Fig.~\ref{fig:powerspectra}(b)
and that energy from that mode is then nonlinearly
coupled into other degrees of freedom.

\onecolumngrid

\section{Additional data sets}
\label{sec:additionaldatasets}

\begin{figure*}[ht]
  \begin{subfigure}[t]{0.48\textwidth}
    \centering
    \includegraphics[width=0.85\linewidth]{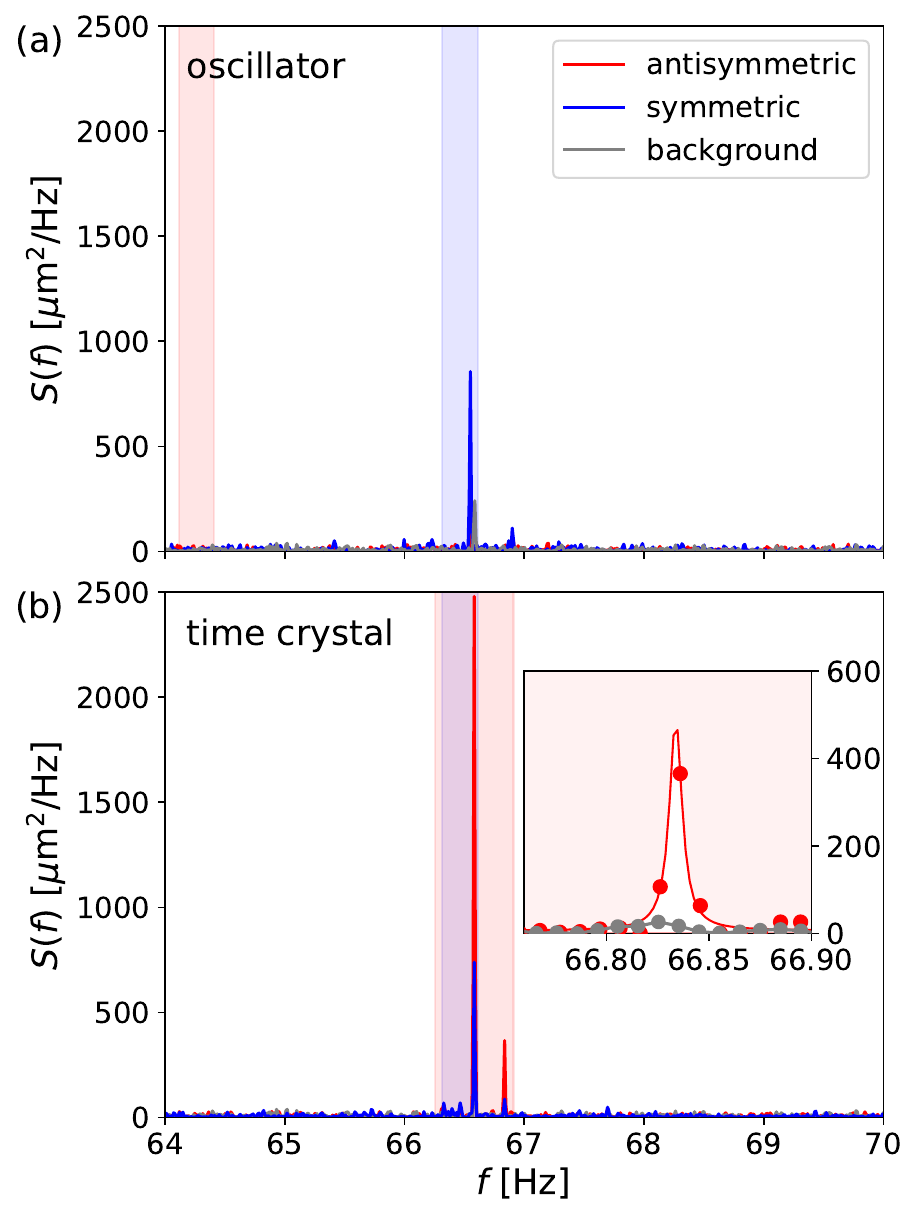}
    \caption{(a) Active oscillator: $ka_1 = \num{0.74(3)}$ and $ka_2 = \num{1.02(2)}$.
      (b) Time crystal: $ka_1 = \num{1.03(5)}$ and $ka_2 = \num{1.22(2)}$.}
    \label{fig:suppdata1}
  \end{subfigure} \hfill
  \begin{subfigure}[t]{0.48\textwidth}
    \centering
    \includegraphics[width=0.85\linewidth]{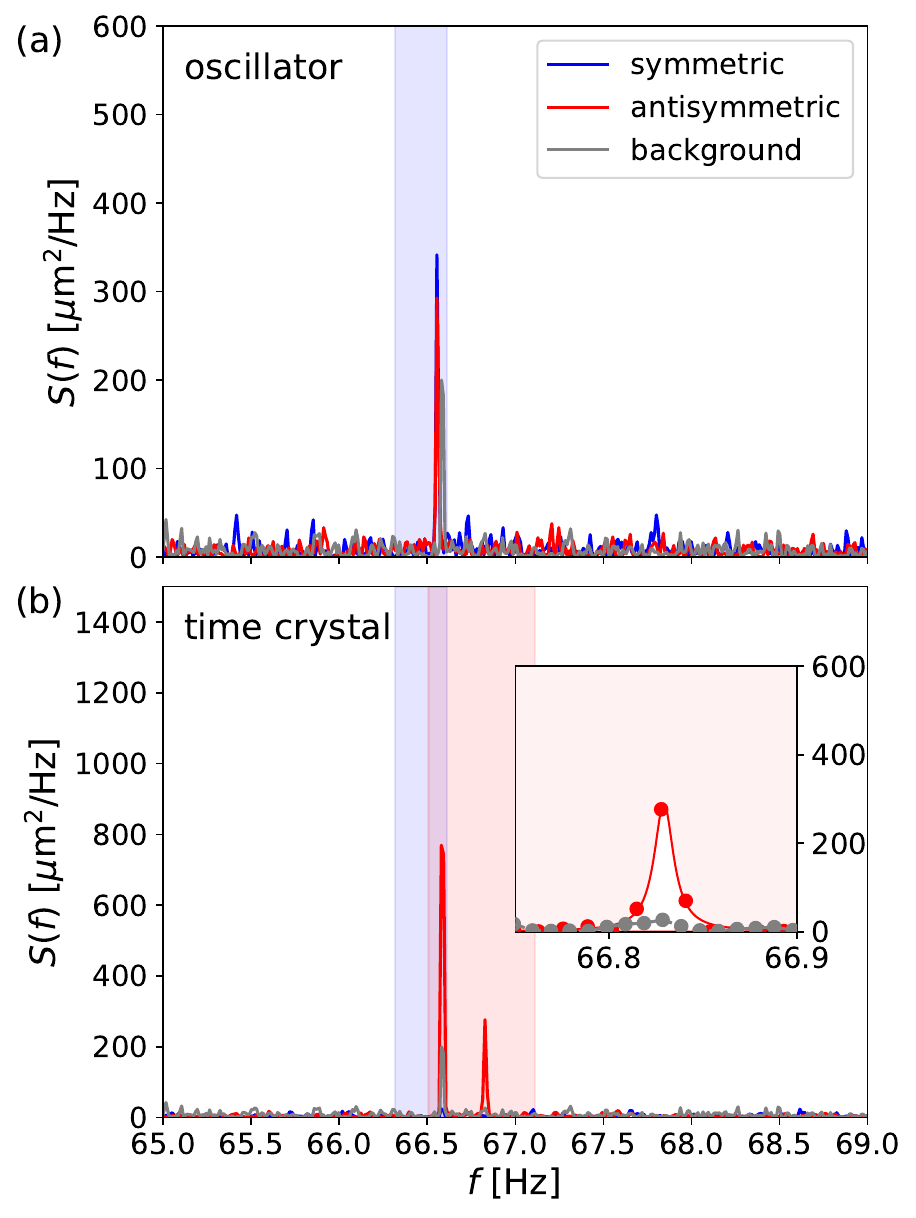}
    \caption{(a) Active oscillator:
      $ka_1 = \num{0.82(2)}$ and $ka_2 = \num{1.02(2)}$.
      (b) Time crystal:
      $ka_1 = \num{1.10(3)}$ and $ka_2 = \num{1.17(4)}$.}
    \label{fig:suppdata2}
  \end{subfigure}

  \medskip

  \begin{subfigure}[t]{0.48\textwidth}
    \centering
    \includegraphics[width=0.85\linewidth]{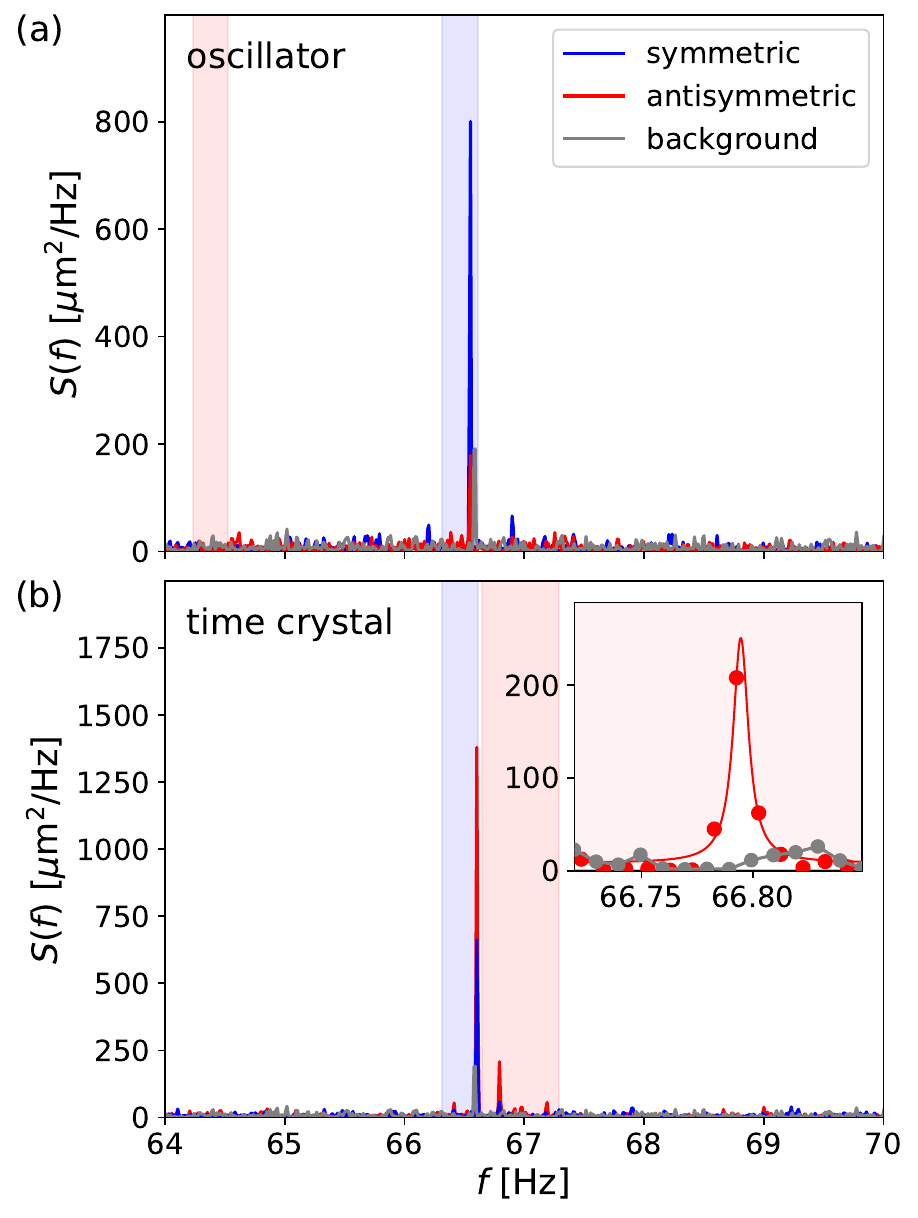}
    \caption{(a) Active oscillator:
      $ka_1 = \num{0.80(2)}$ and $ka_2 = \num{1.02(2)}$.
      (b) Time crystal:
      $ka_1 = \num{1.07(4)}$ and $ka_2 = \num{1.21(3)}$.}
    \label{fig:repeat1}
  \end{subfigure} \hfill
  \begin{subfigure}[t]{0.48\textwidth}
    \centering
    \includegraphics[width=0.85\linewidth]{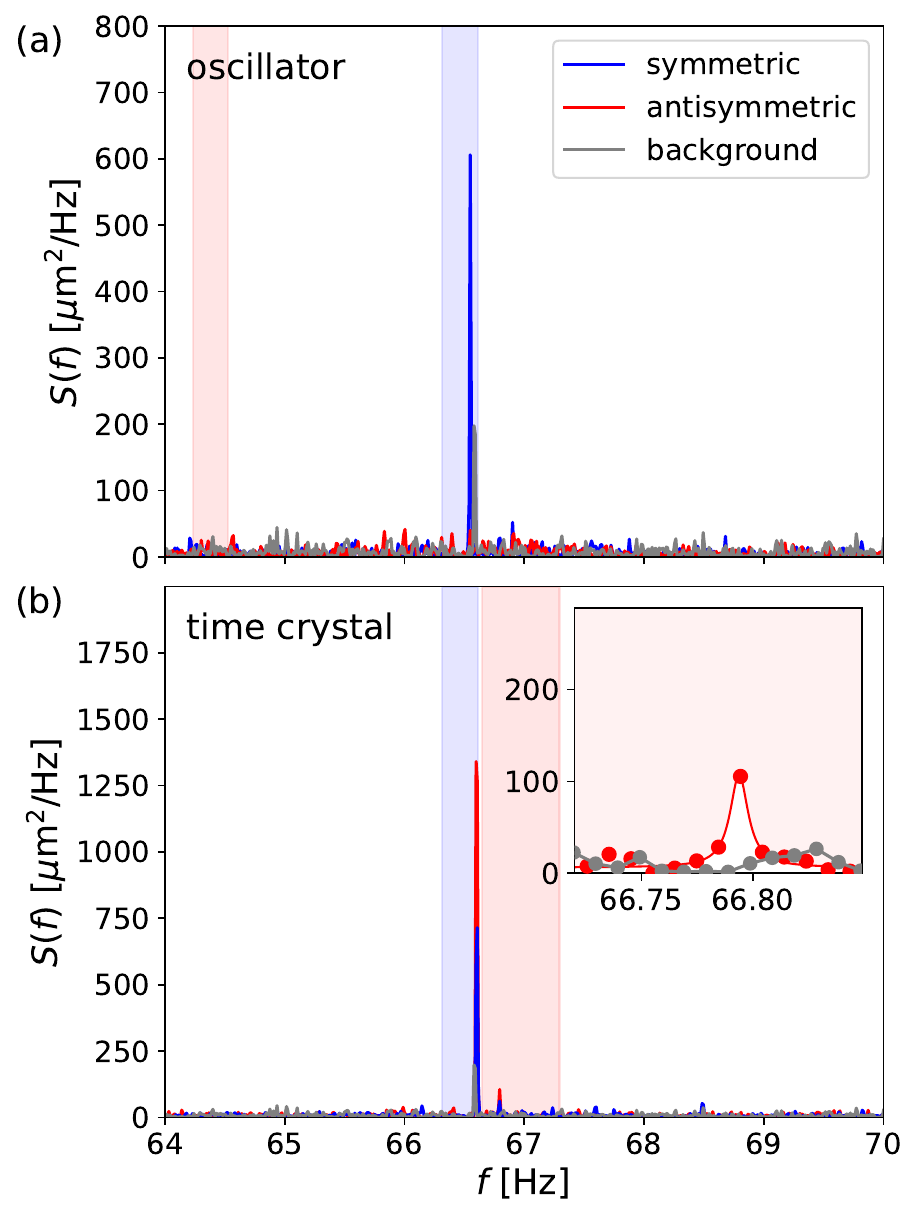}
    \caption{(a) Active oscillator:
      $ka_1 = \num{0.80(2)}$ and $ka_2 = \num{1.02(2)}$.
      (b) Time crystal:
      $ka_1 = \num{1.07(4)}$ and $ka_2 = \num{1.21(3)}$.}
    \label{fig:repeat2}
  \end{subfigure}
  \caption{Power spectral densities for additional data sets.}
  \label{fig:additional}
\end{figure*}

In addition to the two datasets showcased in the main text,
Fig.~\ref{fig:additional} presents power spectral
densities for eight additional datasets, each
obtained from \num{20000}-frame video sequences
acquired at \qty{200}{frames\per\second}.
The four datasets presented in Fig.~\ref{fig:additional}A and Fig.~\ref{fig:additional}B
show results comparable to those in Fig.~3 in the main text but for other pairs of particles.
Agreement between observed and predicted normal-mode
frequencies add support to the claimed predictive
power of the one-dimensional dynamical model
presented in the main text.

The other four datasets displayed in
Figs.~\ref{fig:additional}C and \ref{fig:additional}D
are obtained for the same pairs of particles
used in Fig.~3, but from video sequences obtained
\qty{5}{\minute} later.
These power spectral densities yield normal-mode
frequencies in agreement with
those presented in Fig.~3, supporting the claim
that these systems are in nonequilibrium steady-states.

For each dataset, the power spectral density for the
symmetric mode is plotted in blue and that for the
antisymmetric mode is plotted in red.
Results for the smaller of the two particles
stably trapped in isolation are plotted in gray
and provide an estimate for the measurements'
noise floor.
Colored bands represent the predicted normal-mode
frequencies of the symmetric and antisymmetric
modes based on the particles' measured sizes
and the material properties tabulated in
Table~\ref{tab:materials}.
Insets provide a zoomed-in view of the antisymmetric
mode under conditions that are predicted to support
a time crystal. Solid curves are fits to a
Lorentzian line shape for the peak height, the
noise floor and the persistence time of the
antisymmetric mode.

The symmetric-mode frequency for every pair of
particles is consistent with
the natural frequency, $\Omega_0$, as predicted
in the main text.
The antisymmetric-mode frequency is
predicted to be lower than $\Omega_0$ in
the active-oscillator state, although no significant
spectral peaks are observed within the predicted
range.
This observation is consistent with the prediction
that the antisymmetric mode is highly damped
under conditions that support a steady-state
active oscillator.

The antisymmetric-mode frequency in a steady-state time crystal is predicted to exceed $\Omega_0$.
Spectral peaks in the range of predicted
antisymmetric-mode frequencies are observed
for all pairs of particles whose sizes support
a steady-state time crystal.

All experimental results from the main text and these
additional data sets and their metadats are summarized in Tables \ref{tab:experiments}-\ref{tab:experiment910}.
Data and analysis files for all datasets are available for download at Ref.~\cite{OSF}.

\clearpage

\section{Tables of experimental results}

The videos, trajectory data and analysis code
for the experimental realizations described
in this work are freely available in a public
data repository \cite{OSF}.
Table~\ref{tab:experiments} numbers the
experimental realizations and associates
each video with the associated tracking data
and analysis code.

Tables~\ref{tab:experiment1234}, \ref{tab:experiment5678}, and \ref{tab:experiment910}
summarize the experimental conditions
and analytical outcomes for each of these
experiments.

\begin{table*}[ht]
\centering
\caption{Summary of experiments and associated tracking and analysis files.}
\begin{tabular}{l l l}
\hline
\textbf{Experiment} & \textbf{Data file} & \textbf{Analysis file} \\
\hline
 \# \hspace{40pt}
& Tracking (.csv) and raw (.avi) filestem \hspace{20pt}
& Analysis (.ipynb) filestem \hspace{20pt} \\
\hline
1 & \texttt{oscillationsgreensmall02} & \texttt{powerspectrum\_final}\\
2 & \texttt{oscillationsblueyellowbig03} & \texttt{powerspectrum\_final}\\
3 & \texttt{oscillationsgreengreensmall} & \texttt{powerspectrum\_final02}\\
4 & \texttt{oscillationsorangebluebig03} & \texttt{powerspectrum\_final02}\\
5 & \texttt{oscillationsgreenpeachsmall03} & \texttt{powerspectrum\_final03}\\
6 & \texttt{oscillationsblueredbig} & \texttt{powerspectrum\_final03}\\
7 &
\texttt{oscillationsgreensmall03} & \texttt{powerspectrum\_final\_sc02}\\
8 &
\texttt{oscillationsblueyellowbig02} & \texttt{powerspectrum\_final\_sc02}\\
9 &
\texttt{oscillationsgreensmall} & \texttt{powerspectrum\_final\_sc03}\\
10 &
\texttt{oscillationsblueyellowbig} & \texttt{powerspectrum\_final\_sc03}\\
\hline
\end{tabular}
\label{tab:experiments}
\end{table*}

\begin{table*} [ht]
\centering
\renewcommand{\arraystretch}{1.4}
\caption{Experimental parameters in the oscillator and time-crystal regimes.}
\begin{tabular}{l l c c c c c c}
\hline
\textbf{Parameter}
& \textbf{Description}
& \textbf{Exp.~1}
& \textbf{Exp.~2}
& \textbf{Exp.~3}
& \textbf{Exp.~4}
\\
\hline

$ka_{1}$
& reduced particle 1 radius
& $\num{0.80(2)}$
& $\num{1.07(4)}$
& $\num{0.74(3)}$
& $\num{1.03(5)}$
\\

$ka_{2}$
& reduced particle 2 radius
& $\num{1.02(2)}$
& $\num{1.21(3)}$
& $\num{1.02(2)}$
& $\num{1.22(2)}$
\\

$f_s^{(\text{theory})}$
& symmetric  freq. (theory)
& \qty{66.5(3)}{\hertz}
& \qty{66.5(3)}{\hertz}
& \qty{66.5(3)}{\hertz}
& \qty{66.5(3)}{\hertz}
\\

$f_a^{(\text{theory})}$
& antisymmetric  freq. (theory)
& \qty{64.4(3)}{\hertz}
& \qty{67.0(6)}{\hertz}
& \qty{64.3(3)}{\hertz}
& \qty{66.6(7)}{\hertz}
\\

$f_s^{(\text{data})}$
& symmetric freq. (data)
& \qty{66.5}{\hertz}
& \qty{66.6}{\hertz}
& \qty{66.6}{\hertz}
& \qty{66.6}{\hertz}
\\

$f_a^{(\text{data})}$
& antisymmetric freq. (data)
& n/a
& \qty{66.8}{\hertz}
& n/a
& \qty{66.8}{\hertz}
\\

$t_{c}$
& time crystal coherence time
& n/a
& \qty{102(14)}{\second}
& n/a
& \qty{134(28)}{\second}
\\

SNR
& time crystal signal/noise ratio
& n/a
& $\num{42}$
& n/a
& $\num{71}$
\\

$T$
& length of video
& $\qty{104}{\second}$
& $\qty{108}{\second}$
& $\qty{103}{\second}$
& $\qty{103}{\second}$
\\

Identity
& oscillator (O)/time crystal (TC)
& O
& TC
& O
& TC
\\

Figure
& used in figure panel(s)
& 3(a)
& 3(b)
& 8A(a)
& 8A(b)
\\

\hline
\end{tabular}
\label{tab:experiment1234}
\end{table*}

\begin{table*}[ht]
\centering
\renewcommand{\arraystretch}{1.4}
\caption{Experimental parameters in the oscillator and time-crystal regimes.}
\begin{tabular}{l l c c c c c c}
\hline
\textbf{Parameter}
& \textbf{Description}
& \textbf{Exp.~5}
& \textbf{Exp.~6}
& \textbf{Exp.~7}
& \textbf{Exp.~8}
\\
\hline

$ka_{1}$
& reduced particle 1 radius
& $\num{0.82(2)}$
& $\num{1.10(3)}$
& $\num{0.80(2)}$
& $\num{1.07(4)}$
\\

$ka_{2}$
& reduced particle 2 radius
& $\num{1.02(2)}$
& $\num{1.17(4)}$
& $\num{1.02(2)}$
& $\num{1.21(3)}$
\\

$f_s^{(\text{theory})}$
& symmetric  freq. (theory)
& \qty{66.5(3)}{\hertz}
& \qty{66.5(3)}{\hertz}
& \qty{66.5(3)}{\hertz}
& \qty{66.5(3)}{\hertz}
\\

$f_a^{(\text{theory})}$
& antisymmetric  freq. (theory)
& \qty{64.4(3)}{\hertz}
& \qty{66.8(6)}{\hertz}
& \qty{63.4(1)}{\hertz}
& \qty{67.0(6)}{\hertz}
\\

$f_s^{(\text{data})}$
& symmetric freq. (data)
& \qty{66.6}{\hertz}
& \qty{66.6}{\hertz}
& \qty{66.6}{\hertz}
& \qty{66.6}{\hertz}
\\

$f_a^{(\text{data})}$
& antisymmetric freq. (data)
& n/a
& \qty{66.8}{\hertz}
& n/a
& \qty{66.8}{\hertz}
\\

$t_{c}$
& time crystal coherence time
& n/a
& \qty{80(8)}{\second}
& n/a
& \qty{120(36)}{\second}
\\

SNR
& time crystal signal/noise ratio
& n/a
& $\num{1541}$
& n/a
& $\num{30}$
\\

$T$
& length of video
& $\qty{102}{\second}$
& $\qty{77}{\second}$
& $\qty{110}{\second}$
& $\qty{101}{\second}$
\\

Identity
& oscillator (O)/time crystal (TC)
& O
& TC
& O
& TC
\\

Figure
& used in figure panel(s)
& 8B(a)
& 8B(b)
& 8C(a)
& 8C(b)
\\

\hline
\end{tabular}
\label{tab:experiment5678}
\end{table*}

\begin{table*}[ht]
\centering
\renewcommand{\arraystretch}{1.4}
\caption{Experimental parameters in the oscillator and time-crystal regimes.}
\begin{tabular}{l l c c c c c c}
\hline
\textbf{Parameter}
& \textbf{Description}
& \textbf{Exp.~9}
& \textbf{Exp.~10}
\\
\hline

$ka_{1}$
& reduced particle 1 radius
& $\num{0.80(2)}$
& $\num{1.07(4)}$
\\

$ka_{2}$
& reduced particle 2 radius
& $\num{1.02(2)}$
& $\num{1.21(3)}$
\\

$f_s^{(\text{theory})}$
& symmetric  freq. (theory)
& \qty{66.5(3)}{\hertz}
& \qty{66.5(3)}{\hertz}
\\

$f_a^{(\text{theory})}$
& antisymmetric  freq. (theory)
& \qty{64.4(3)}{\hertz}
& \qty{67.0(7)}{\hertz}
\\

$f_s^{(\text{data})}$
& symmetric freq. (data)
& \qty{66.6}{\hertz}
& \qty{66.6}{\hertz}
\\

$f_a^{(\text{data})}$
& antisymmetric freq. (data)
& n/a
& \qty{66.8}{\hertz}
\\

$t_{c}$
& time crystal coherence time
& n/a
& \qty{101(16)}{\second}
\\

SNR
& time crystal signal/noise ratio
& n/a
& $\num{17}$
\\

$T$
& length of video
& $\qty{105}{\second}$
& $\qty{102}{\second}$
\\

Identity
& oscillator (O)/time crystal (TC)
& O
& TC
\\

Figure
& used in figure panel(s)
& 8D(a)
& 8D(b)
\\

\hline
\end{tabular}
\label{tab:experiment910}
\end{table*}

\clearpage

\twocolumngrid

%

\end{document}